\documentclass[twocolumn,preprintnumbers,aps,superscriptaddress,nofootinbib]{revtex4-2}
\usepackage{amsmath,amssymb,nccmath,graphicx,color,hyperref}
\usepackage{scalefnt}
\definecolor{darkblue}{rgb}{0.0,0.0,0.7}
\definecolor{nicered}{rgb}{0.7,0.1,0.1}
\definecolor{nicegreen}{rgb}{0.0,0.4,0.0}
\hypersetup{colorlinks,citecolor=nicegreen,linkcolor=nicered,urlcolor=nicered}
\graphicspath{ {fig/} }
\makeatletter
\newcommand{\oset}[3][0ex]{%
  \mathrel{\mathop{#3}\limits^{
    \vbox to#1{\kern-2\ex@
    \hbox{$\scriptstyle#2$}\vss}}}}
\makeatother
\newcommand{\ep}{\epsilon}
\newcommand{\CS}{\textcolor{darkblue}}
\newcommand{\pole}[1]{\textcolor{darkblue}{#1}}
\newcommand{\mint}[1]{\scalebox{0.85}{$#1$\,}}
\newcommand{\Graph}[2]{\vcenter{\hbox{\includegraphics[scale=#1]{#2}}}}
\newcommand{\FDiag}[2]{
\begin{minipage}{0.12\textwidth}
$\includegraphics[width=\textwidth]{#2}\hspace*{-13ex}\raisebox{-1.5ex}{\CS{#1}}\hfill$
\end{minipage}
\hspace*{-2ex}
}

\allowdisplaybreaks[4]

\begin{document}

\preprint{MSUHEP-22-003, P3H-22-014, TTP22-008}

\title{Quark and gluon form factors in four-loop QCD}

\author{Roman N. Lee}
\affiliation{Budker Institute of Nuclear Physics, 630090 Novosibirsk, Russia}

\author{Andreas von Manteuffel}
\affiliation{Department of Physics and Astronomy, Michigan State University,
East Lansing, Michigan 48824, USA}

\author{Robert M. Schabinger}
\affiliation{Department of Physics and Astronomy, Michigan State University,
East Lansing, Michigan 48824, USA}

\author{Alexander V. Smirnov}
\affiliation{Research Computing Center, Moscow State University,
119991, Moscow, Russia}
\affiliation{Moscow Center for Fundamental and Applied Mathematics,
119992, Moscow, Russia}

\author{Vladimir A. Smirnov}
\affiliation{Skobeltsyn Institute of Nuclear Physics of Moscow State University,
119991, Moscow, Russia}
\affiliation{Moscow Center for Fundamental and Applied Mathematics,
119992, Moscow, Russia}

\author{Matthias Steinhauser}
\affiliation{Institut f{\"u}r Theoretische Teilchenphysik,
Karlsruhe Institute of Technology (KIT),
76128 Karlsruhe, Germany}

\begin{abstract}
  \noindent
  We compute the photon-quark and Higgs-gluon form factors to four-loop order
  within massless perturbative Quantum Chromodynamics.
  Our results constitute ready-to-use building blocks for N${}^4$LO cross
  sections for Drell-Yan processes and gluon-fusion Higgs boson production at the LHC.
  We present complete analytic expressions for both form factors
  and show several of the most complicated master integrals.
\end{abstract}

\maketitle



{\bf Introduction.}
A number of experimental results obtained at the Large Hadron
Collider (LHC) at CERN have reached a precision below the percent level,
often superseding the original expectations.  A fundamental ingredient to
the successful interpretation of precise data is the computation of
higher order quantum corrections, most
importantly those stemming from the strong interaction. In many cases
next-to-next-to-leading order (NNLO) corrections have become standard. In fact,
nowadays $2\to2$ scattering processes are routinely computed at this order,
in some cases even taking into account massive particles in the loops.
Also for $2\to3$ processes more and more results become available
(see, \textit{e.g.}, Refs.~\cite{Chawdhry:2019bji,Kallweit:2020gcp,Czakon:2021mjy,Agarwal:2021vdh,Badger:2021imn,Badger:2021nhg,Abreu:2021asb}).

There are a few processes which are known to third order, or N$^3$LO,
in perturbative Quantum Chromodynamics (QCD).
Among them are the Drell-Yan production of $W$ and $Z$ bosons~\cite{Duhr:2020sdp,Duhr:2021vwj}
as well as Higgs boson production in gluon fusion
in the infinite top-mass limit~\cite{Anastasiou:2015vya,Mistlberger:2018etf}
at the LHC.
In the latter case the high-order corrections are particularly important
due to the slow convergence of the perturbative series.
Similar observations have been made for the threshold
production cross section of the top quark pairs in electron positron annihilation,
where third-order corrections are necessary to obtain theory uncertainties
of a few percent~\cite{Beneke:2015kwa}.
For more generic $2\to2$ processes like dijet production, virtual corrections
at third order QCD became available only recently,
(see, \textit{e.g.}, Refs.~\cite{Caola:2021izf,Henn:2020lye}),
providing first ingredients to such N$^3$LO cross sections.

In the coming years the precise determination of the Higgs boson properties
will be one of the central topics at the LHC. In this context it is important
to improve the precision for the production cross section, both
experimentally and from the theory side.  First steps towards the N$^4$LO
corrections of the Higgs boson production cross section have been undertaken
in Ref.~\cite{Das:2020adl}. In this Letter we provide the first ready-to-use
ingredient to the N$^4$LO cross section for $gg\to H + X$
by presenting the virtual
corrections to the effective Higgs-gluon vertex up to four-loop order.
Similarly, we provide the four-loop corrections to the photon-quark vertex
which are a building block of the N$^4$LO corrections to the process $q\bar{q} \to Z/W$.
Historically, also at N$^3$LO the purely virtual corrections were the first
building blocks to become available with the calculation of the three-loop form
factors more than a decade ago~\cite{Baikov:2009bg,Lee:2010cga,Gehrmann:2010ue}.
Subsequently, the real-radiation contributions have been added
step-by-step until first results for the Higgs production cross section became available~\cite{Anastasiou:2013srw,Anastasiou:2014vaa,Duhr:2013msa,Kilgore:2013gba,Li:2013lsa,Li:2014bfa,Li:2014afw}.

The relevant form factors for the $q\bar{q}\gamma^\ast$ and
$ggH$ vertex functions $\Gamma^{\mu}_q$ and $\Gamma^{\mu\nu}_g$, respectively,
are given by the projections
\begin{align}
\label{eq::FFq}
  F_q(q^2) &= -\frac{1}{4(1-\epsilon)q^2}
  \mbox{Tr}\left( q_2\!\!\!\!\!/\,\,\, \Gamma^\mu q_1\!\!\!\!\!/\,\,\,
    \gamma_\mu\right)
  \,,
 \\
\label{eq::FFg}
  F_g(q^2) &=
  \frac{\left(q_1\cdot q_2\,\,
      g_{\mu\nu}-q_{1,\mu}\,q_{2,\nu}-q_{1,\nu}\,q_{2,\mu}\right)}
  {2(1-\epsilon)}
  \Gamma^{\mu\nu}_g
  \,.
\end{align}
Here, the overall normalization is chosen such that both form factors are one at
leading order.
We employ conventional dimensional regularization and use $\epsilon=(4-d)/2$,
where $d$ is the space-time dimension.
The outgoing momentum of the photon (Higgs) is $q=q_1+q_2$,
where $q_1$ and $q_2$ are the incoming momenta of the quark and anti-quark
(gluons) for $F_q$ ($F_g$), and we have $q_1^2=q_2^2=0$ and $q^2\not=0$.

\begin{figure*}[t]
\begin{center}
 \raisebox{1ex}
 {$F_q:\!\!$}
 \FDiag{$C_F^4$}{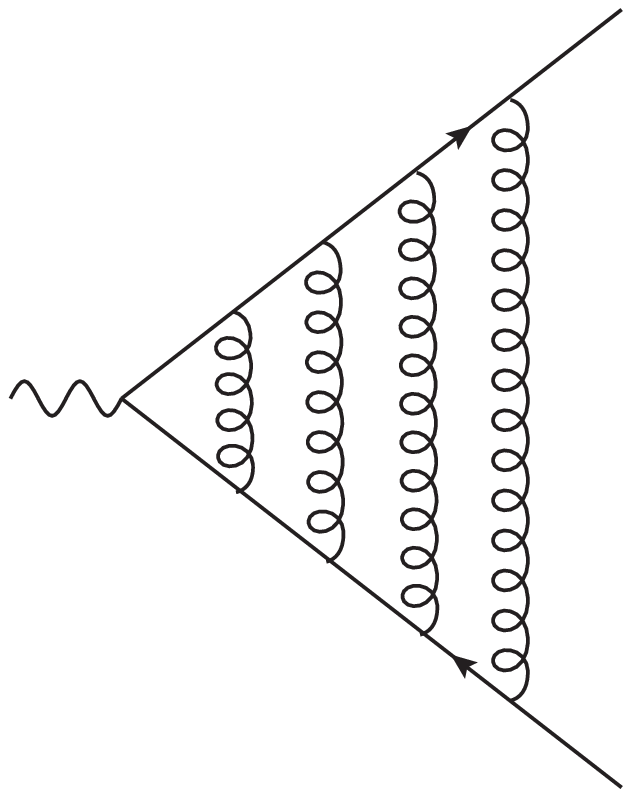}
 \FDiag{$C_F^3 C_A$}{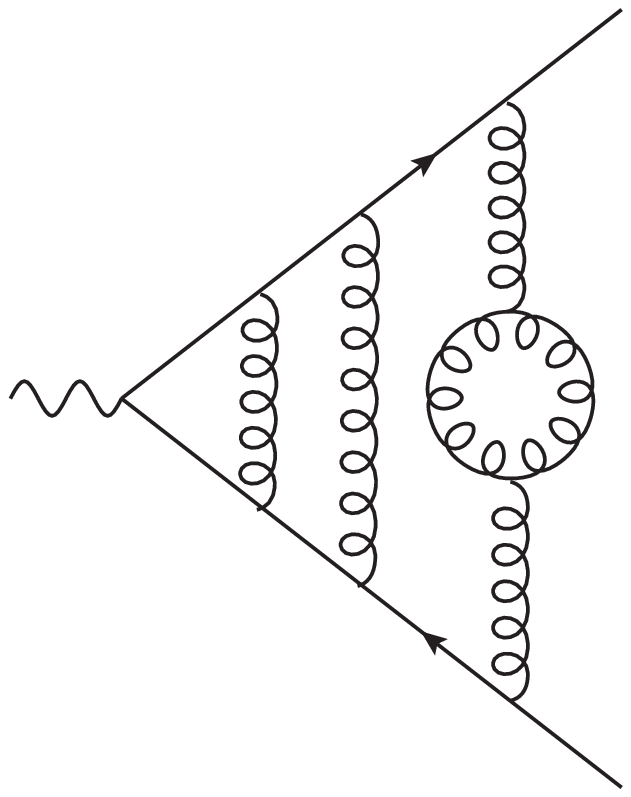}
 \FDiag{$C_F^2 C_A^2$}{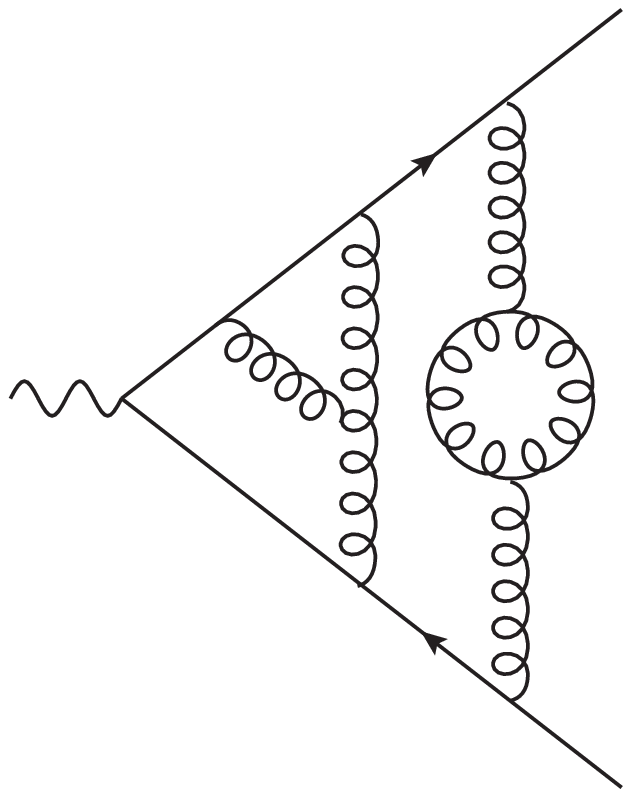}
 \FDiag{$C_F C_A^3$}{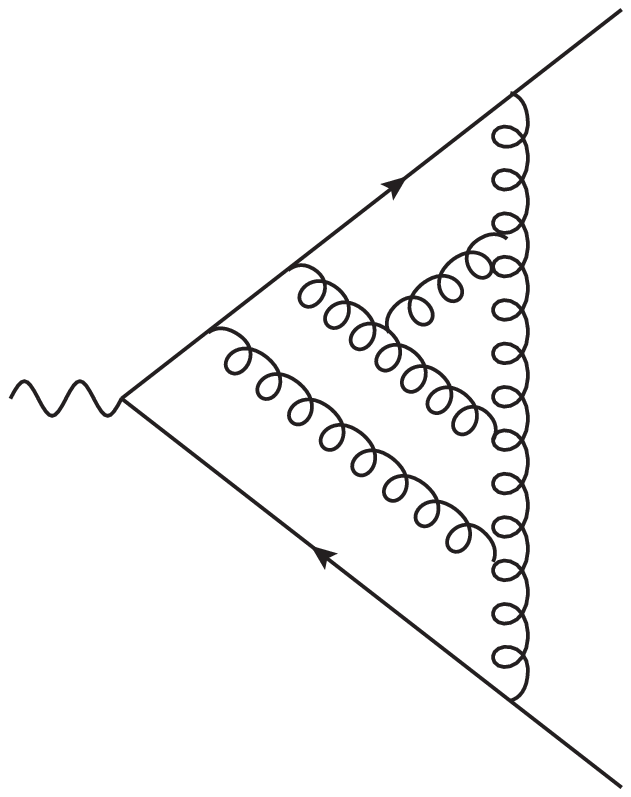}
 \FDiag{$d_F^{abcd}d_A^{abcd}$}{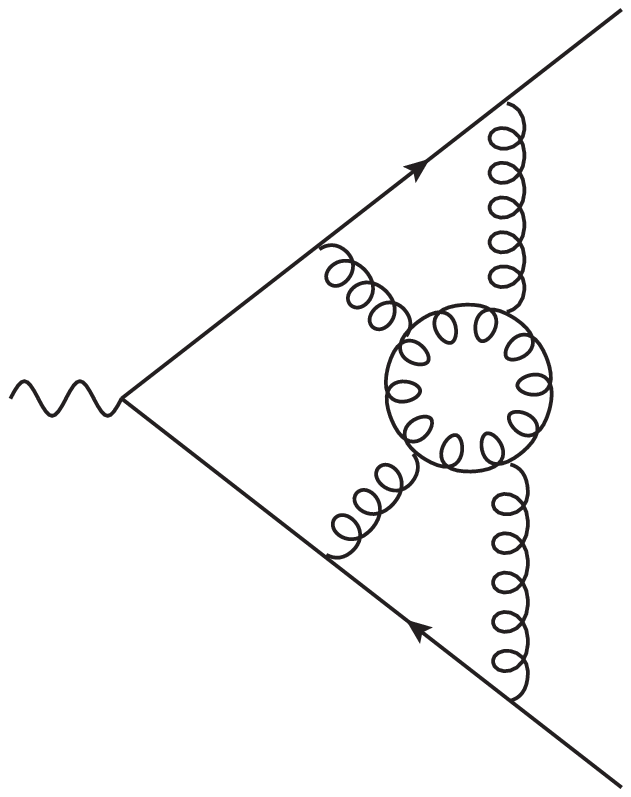}
 \hfill
 \raisebox{1ex}
 {$F_g:\!\!$}
 \FDiag{$C_A^4$}{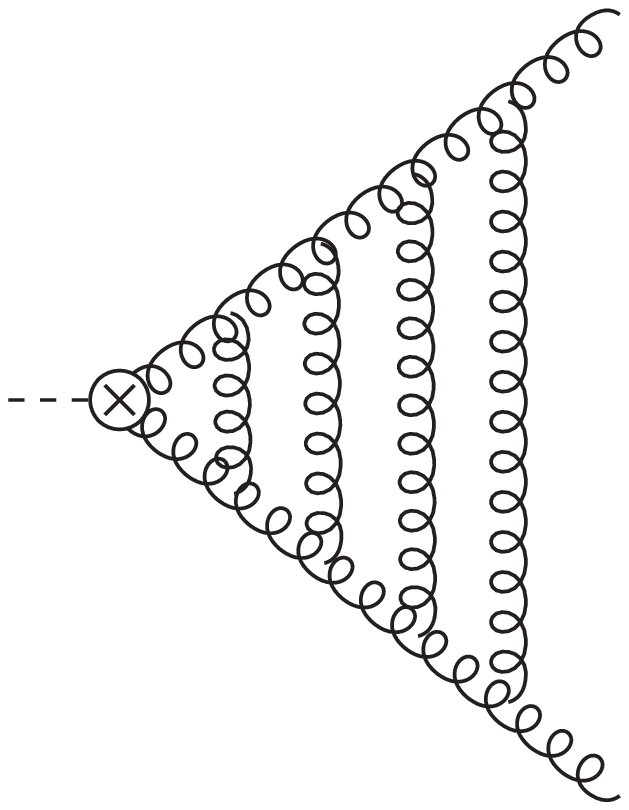}
 \FDiag{$d_A^{abcd}d_A^{abcd}$}{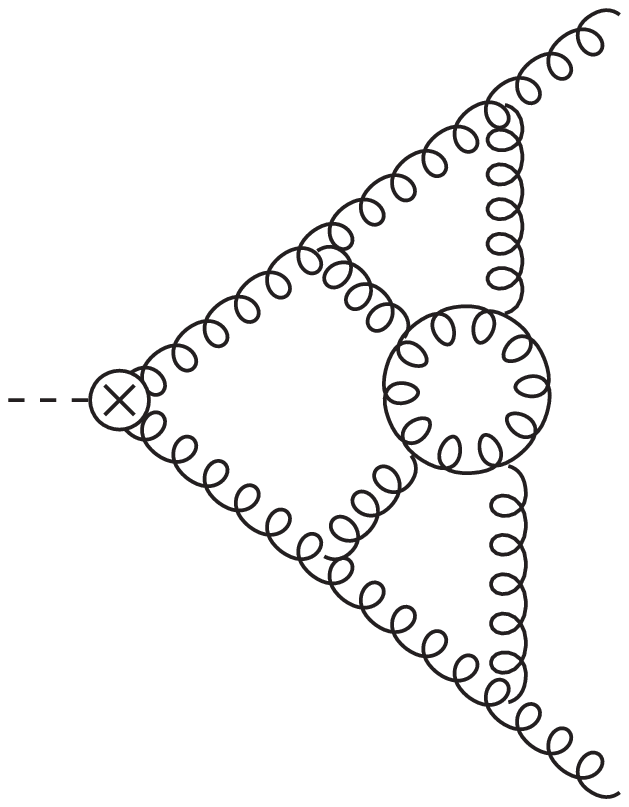}
\end{center}
\caption{\label{fig::diags}Sample Feynman diagrams and color factors
  for the non-fermionic contributions to $F_g$ and $F_q$
  at four-loop order. Straight and curly lines denote
  quarks and gluons, respectively.
  Both planar and non-planar diagrams contribute.
}
\end{figure*}
The two- and three-loop QCD corrections to $F_q$ and $F_g$ are available from
Refs.~\cite{Kramer:1986sg,Matsuura:1987wt,Matsuura:1988sm,Gehrmann:2005pd,Harlander:2000mg,Baikov:2009bg,Lee:2010ik,Baikov:2009bg,Gehrmann:2010ue,Gehrmann:2010tu,vonManteuffel:2015gxa,Heinrich:2009be},
At four loops, only partial results have been obtained so far.
In Fig.~\ref{fig::diags} we show sample Feynman diagrams for the purely gluonic
corrections to $F_q$ and $F_g$ in four-loop QCD; sample diagrams for the fermionic part can be found in Fig.~1 of Ref.~\cite{Lee:2021uqq}.
Altogether 5728 and 43220 Feynman diagrams contribute to the quark and gluon form factor at this perturbative order, respectively.

The results, which are presented in this Letter, finalize a long-running
effort to compute QCD form factors to four loops.  First results have been
obtained in 2016~\cite{Henn:2016men,Lee:2016ixa} where planar diagrams for
$F_q$ have been presented in the large-$N_c$ limit.
Fermionic corrections with two closed fermion bubbles are
available from~\cite{Lee:2017mip} and the complete contribution from color
structure $(d_F^{abcd})^2$ has been computed
in~\cite{Lee:2019zop,vonManteuffel:2020vjv}.  For $F_q$ and $F_g$, all
corrections with three or two closed fermion loops have been calculated in
\cite{vonManteuffel:2016xki,vonManteuffel:2019wbj}, respectively, including
also the singlet contributions.
The complete set of poles of $F_q$ and $F_g$ in the dimensional regulator
has been obtained through direct diagrammatic evaluation
in~\cite{Agarwal:2021zft}.
Finally, the complete fermionic corrections to $F_q$ and $F_g$
have been computed in Ref.~\cite{Lee:2021uqq}.


\bigskip
{\bf Calculation.} The calculation of the four-loop form factors
presents two major challenges.
The first one is connected to a minimal representation of the form factors.
After generating the Feynman diagrams with \texttt{Qgraf}~\cite{Nogueira:1991ex},
we apply the projectors and perform the numerator and color algebra
with \texttt{Form\;4}~\cite{Kuipers:2012rf} and \texttt{Color.h}~\cite{vanRitbergen:1998pn}.
In this way, we can write the form factors as a linear combination
of a large number of scalar Feynman integrals, each belonging to
one of 100 twelve-line top-level topologies or a subtopology
thereof.
Fixing the twelve propagators and six irreducible numerators
of its top-level topology, a scalar integral can be described
by eighteen integers indicating the exponents of the propagators
and numerators.
By choosing the irreducible numerators as suitably defined inverse
propagators, all top-level topologies can be described in terms of the ten complete
sets of denominators described in \cite{Lee:2021lkc}.
Integration-by-parts (IBP)
reductions~\cite{Tkachov:1981wb,Chetyrkin:1981qh,Laporta:2001dd} systematically
establish linear relations between the integrals, allowing us to
express the form factors as a linear combination
of a minimal set of so-called master integrals.
For our calculation we use the setup described in~\cite{vonManteuffel:2020vjv}
based on the program \texttt{Reduze\;2}~\cite{vonManteuffel:2012np} and
the in-house code \texttt{Finred}, employing techniques from~\cite{vonManteuffel:2014ixa,Gluza:2010ws,Larsen:2015ped,Boehm:2017wjc,Lee:2014tja,Bitoun:2017nre,Agarwal:2020dye}.

The second challenge is the computation of the master integrals.  Here we
follow two complementary approaches.  The first one is based on the
construction of finite master
integrals~\cite{vonManteuffel:2014qoa,vonManteuffel:2015gxa,Schabinger:2018dyi},
in $d_0-2\epsilon$ dimensions where $d_0=4,6,\ldots$.
Provided a linearly reducible~\cite{Brown:2008um,Brown:2009ta}
Feynman parametric representation can be found,
the $\epsilon$ expansions of such master integrals may be computed analytically using the program {\tt HyperInt}~\cite{Panzer:2014caa}.
The dimensionally shifted integrals can be related to master integrals
in $4-2\epsilon$ dimensions using IBP relations derived 
with first- and second-order annihilators in the Lee-Pomeransky
representation~\cite{Lee:2013hzt}.
We wish to point out that in this approach, the integration can be performed
at the level of individual integrals.
In practice, evaluating higher orders of the $\epsilon$ expansion gets
ever more demanding due to the rise in algebraic complexity.
To determine the form factors $F_q$ and $F_g$, we computed a number of
integrals to transcendental weight eight in this approach, including
computationally demanding non-planar integrals with twelve different propagators.
For one such irreducible topology with a single twelve-line master integral
we find
\begin{widetext}
\begin{align}
\label{eq:resE112471}
&\Graph{1}{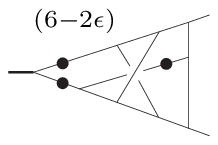} = \Big(\!
  - \mfrac{119}{48}\zeta_7
  - \mfrac{5}{6}\zeta_5 \zeta_2
  - \mfrac{53}{10}\zeta_3 \zeta_2^2
  + \mint{3}\zeta_3^2
  + \mfrac{79}{42}\zeta_2^3
  + \mfrac{25}{6}\zeta_5
  - \mfrac{5}{3}\zeta_3 \zeta_2
  + \mfrac{1}{15}\zeta_2^2
  + \mint{2}\zeta_3
 \Big)+ \pole{\ep} \Big(\!
  - \mfrac{991}{30}\zeta_{5,3}
  - \mfrac{323}{2}\zeta_5 \zeta_3
\notag\\ & \qquad
  - \mfrac{81}{2}\zeta_3^2 \zeta_2
  + \mfrac{127223}{31500}\zeta_2^4
  - \mfrac{2827}{24}\zeta_7
  + \mfrac{73}{6}\zeta_5 \zeta_2
  - \mint{14}\zeta_3 \zeta_2^2
  + \mfrac{41}{3}\zeta_3^2
  + \mfrac{1696}{315}\zeta_2^3
  + \mfrac{401}{3}\zeta_5
  + \mfrac{206}{3}\zeta_3 \zeta_2
  + \mfrac{23}{15}\zeta_2^2
  + \mint{14}\zeta_3
 \Big)
 + \mathcal{O}(\pole{\ep^2})
\end{align}
\end{widetext}
in the conventions of Ref.~\cite{vonManteuffel:2019gpr}.
In particular, the integral is defined in $6-2\epsilon$
and each dot indicates a squared propagator.
We would like to mention that no integral in this topology was needed
for the calculation of the $\mathcal{N}=4$ Sudakov form factor~\cite{Lee:2021lkc}.
Our result above is expressed in terms of regular zeta values,
$\zeta_n$ $(n=2,\ldots,7)$, and
\begin{equation}
    \zeta_{5,3} = \sum_{m=1}^\infty \sum_{n=1}^{m-1} \frac{1}{m^5 n^3} \approx 0.0377076729848
\end{equation} 
is the only multiple zeta value involved.

Our second approach for computing master integrals is the method of differential
equations based on ``canonical'' bases~\cite{Kotikov:2010gf,Henn:2013pwa}. Since our master integrals only depend on one kinematic parameter, $q^2$, we have to introduce a second mass scale such that non-trivial
differential equations can be established. With canonical bases, this idea was first implemented in~\cite{Henn:2013nsa}.
For our application it is advantageous to make one of the
massless external legs massive. Choosing $q_1^2\not=0$ has the advantage that
the boundary conditions can be fixed for $q_1^2=q^2$ since then the vertex
integrals turn into two-point integrals, which are well-studied in the
literature~\cite{Baikov:2010hf,Lee:2011jt}.
The differential equations are used
to transport the information to $q_1^2=0$, which corresponds to the vertex
diagrams we are interested in. 
To construct canonical bases we apply the algorithm of Ref.~\cite{Lee:2014ioa} implemented in~\cite{Lee:2020zfb}.
Details of this approach can, \textit{e.g.}, be found in Refs.~\cite{Lee:2019zop}. When constructing canonical bases, we also need IBP reduction to master integrals. Here, we apply {\tt FIRE}~\cite{Smirnov:2019qkx} for this.

For one of the most complicated twelve-line topologies which did not enter
the $\mathcal{N}=4$ Sudakov form factor we obtain for its two master integrals
\begin{widetext}
\begin{align}
&\Graph{.15}{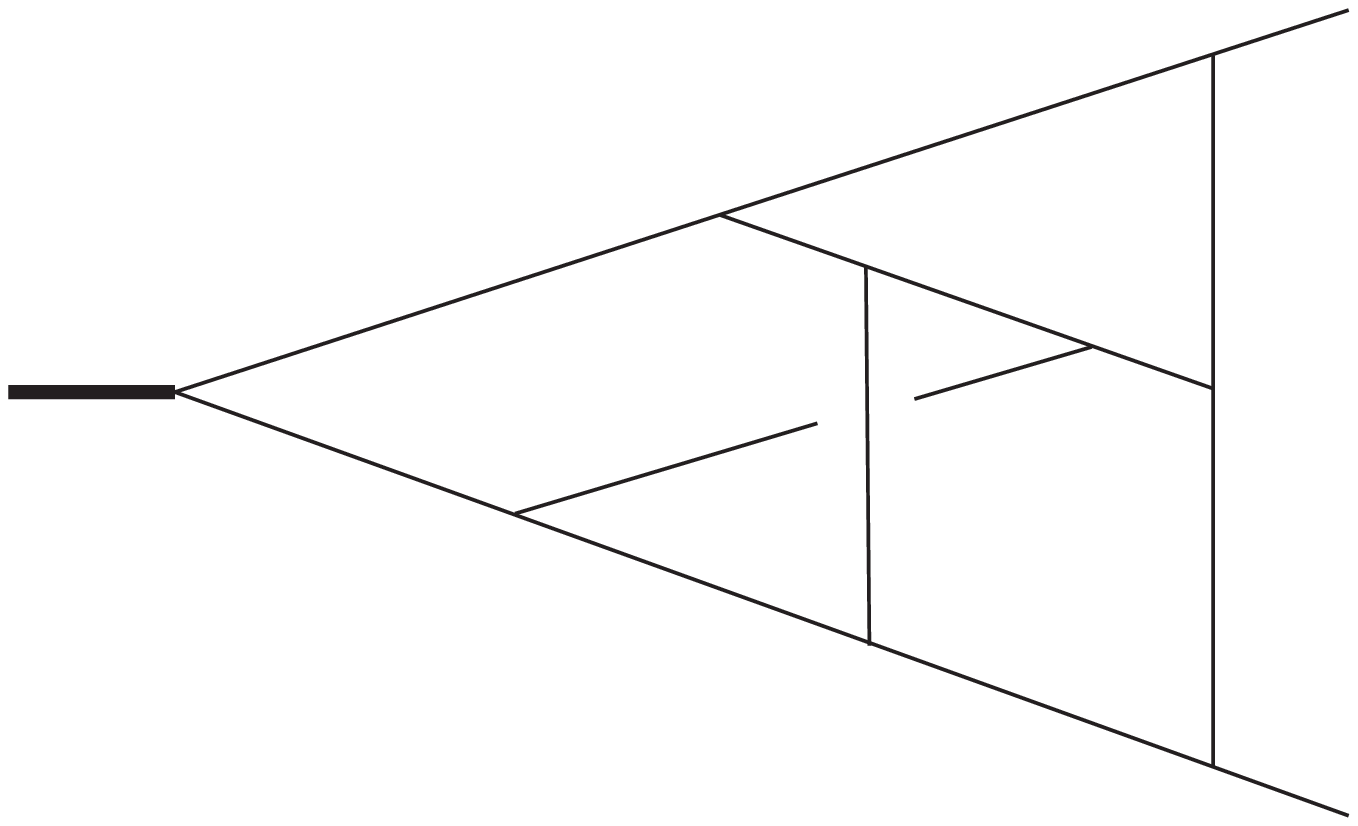} = \pole{\frac{1}{\ep^8}} \Big(
   \mfrac{1}{72}
 \Big)+ \pole{\frac{1}{\ep^7}} \Big(
   \mfrac{17}{96}
 \Big)+ \pole{\frac{1}{\ep^6}} \Big(\!
  - \mfrac{19}{72}\zeta_2
  + \mfrac{5}{36}
 \Big)+ \pole{\frac{1}{\ep^5}} \Big(\!
  - \mfrac{101}{36}\zeta_3
  - \mfrac{719}{288}\zeta_2
  - \mfrac{511}{288}
 \Big)+ \pole{\frac{1}{\ep^4}} \Big(\!
  - \mfrac{1967}{360}\zeta_2^2
  - \mfrac{4123}{144}\zeta_3
\notag\\ & \qquad
  + \mfrac{11}{96}\zeta_2
  + \mfrac{2113}{288}
 \Big)+ \pole{\frac{1}{\ep^3}} \Big(\!
  - \mfrac{697}{8}\zeta_5
  + \mfrac{235}{9}\zeta_3 \zeta_2
  - \mfrac{182903}{2880}\zeta_2^2
  - \mfrac{29}{24}\zeta_3
  + \mfrac{1637}{288}\zeta_2
  - \mfrac{1477}{72}
 \Big)+ \pole{\frac{1}{\ep^2}} \Big(
   \mfrac{27083}{144}\zeta_3^2
  - \mfrac{130951}{2520}\zeta_2^3
\notag\\ & \qquad
  - \mfrac{12907}{16}\zeta_5
  + \mfrac{16883}{144}\zeta_3 \zeta_2
  + \mfrac{2101}{320}\zeta_2^2
  + \mfrac{5527}{36}\zeta_3
  + \mfrac{2899}{144}\zeta_2
  + \mfrac{4223}{144}
 \Big)+ \pole{\frac{1}{\ep}} \Big(\!
  - \mfrac{916123}{384}\zeta_7
  + \mfrac{16565}{48}\zeta_5 \zeta_2
  + \mfrac{727901}{720}\zeta_3 \zeta_2^2
  + \mfrac{26927}{18}\zeta_3^2
\notag\\ & \qquad
  - \mfrac{13321753}{20160}\zeta_2^3
  + \mfrac{2795}{12}\zeta_5
  + \mfrac{147}{4}\zeta_3 \zeta_2
  + \mfrac{979013}{2880}\zeta_2^2
  - \mfrac{77021}{144}\zeta_3
  - \mfrac{5771}{16}\zeta_2
  + \mfrac{13315}{144}
 \Big)
+ \pole{} \Big(
   \mfrac{110723}{120}\zeta_{5,3}
  + \mfrac{104859}{8}\zeta_5 \zeta_3
  - \mfrac{1123}{9}\zeta_3^2 \zeta_2
\notag\\ & \qquad
  - \mfrac{6055979}{21000}\zeta_2^4
  - \mfrac{5285507}{384}\zeta_7
  - \mfrac{1373}{4}\zeta_5 \zeta_2
  + \mfrac{11089517}{1440}\zeta_3 \zeta_2^2
  + \mfrac{88945}{96}\zeta_3^2
  + \mfrac{1391189}{2240}\zeta_2^3
  + \mfrac{20717}{6}\zeta_5
  - \mfrac{118961}{144}\zeta_3 \zeta_2
  - \mfrac{1021111}{720}\zeta_2^2
\notag\\ & \qquad
  + \mfrac{1701}{4}\zeta_3
  + \mfrac{358103}{144}\zeta_2
  - \mfrac{128443}{144}
 \Big)
 + \mathcal{O}(\pole{\ep}) \\
&\Graph{.15}{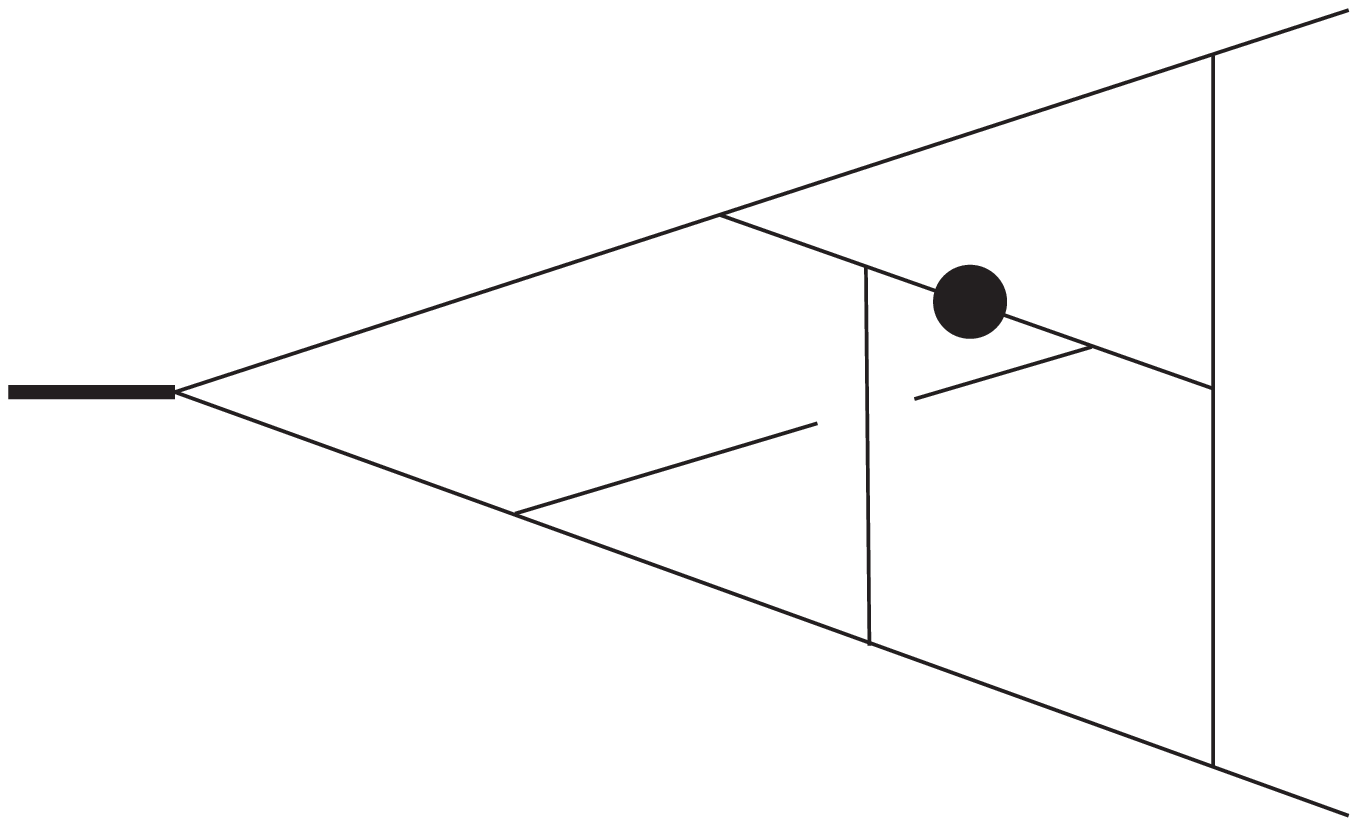} = \pole{\frac{1}{\ep^7}} \Big(\!
  - \mfrac{5}{216}
 \Big)+ \pole{\frac{1}{\ep^6}} \Big(\!
  - \mfrac{13}{36}\zeta_2
  - \mfrac{6157}{2592}
 \Big)+ \pole{\frac{1}{\ep^5}} \Big(\!
  - \mfrac{43}{36}\zeta_3
  - \mfrac{1661}{432}\zeta_2
  - \mfrac{267889}{15552}
 \Big)+ \pole{\frac{1}{\ep^4}} \Big(
   \mfrac{2621}{360}\zeta_2^2
  - \mfrac{536}{27}\zeta_3
  + \mfrac{77653}{2592}\zeta_2
\notag\\ & \qquad
  + \mfrac{3522299}{93312}
 \Big)+ \pole{\frac{1}{\ep^3}} \Big(
   \mfrac{173}{3}\zeta_5
  + \mfrac{722}{9}\zeta_3 \zeta_2
  + \mfrac{235093}{4320}\zeta_2^2
  + \mfrac{806659}{2592}\zeta_3
  + \mfrac{4152625}{15552}\zeta_2
  + \mfrac{1129069}{17496}
 \Big)+ \pole{\frac{1}{\ep^2}} \Big(
   \mfrac{12499}{36}\zeta_3^2
  + \mfrac{6801}{40}\zeta_2^3
\notag\\ & \qquad
  + \mfrac{34745}{72}\zeta_5
  + \mfrac{23542}{27}\zeta_3 \zeta_2
  + \mfrac{287881}{405}\zeta_2^2
  + \mfrac{5714285}{1944}\zeta_3
  + \mfrac{805613}{46656}\zeta_2
  - \mfrac{2167037021}{1679616}
 \Big)+ \pole{\frac{1}{\ep}} \Big(
   \mfrac{137837}{96}\zeta_7
  + \mfrac{5739}{4}\zeta_5 \zeta_2
  + \mfrac{26929}{180}\zeta_3 \zeta_2^2
\notag\\ & \qquad
  + \mfrac{2608907}{432}\zeta_3^2
  + \mfrac{16974331}{10080}\zeta_2^3
  + \mfrac{1920841}{288}\zeta_5
  + \mfrac{579521}{1296}\zeta_3 \zeta_2
  + \mfrac{151664975}{31104}\zeta_2^2
  - \mfrac{210823813}{46656}\zeta_3
  - \mfrac{206363129}{139968}\zeta_2
  + \mfrac{84786801307}{10077696}
 \Big)
 \notag\\ & \qquad
 + \pole{} \Big(\!
  - \mfrac{221827}{60}\zeta_{5,3}
  - \mfrac{1225}{3}\zeta_5 \zeta_3
  - \mfrac{40667}{9}\zeta_3^2 \zeta_2
  + \mfrac{47962441}{28000}\zeta_2^4
  + \mfrac{1542237}{32}\zeta_7
  + \mfrac{114838}{9}\zeta_5 \zeta_2
  + \mfrac{9353537}{2160}\zeta_3 \zeta_2^2
  + \mfrac{7987499}{2592}\zeta_3^2
\notag\\ & \qquad
  + \mfrac{496790909}{90720}\zeta_2^3
  + \mfrac{4253033}{108}\zeta_5
  - \mfrac{127736653}{7776}\zeta_3 \zeta_2
  - \mfrac{8249965}{11664}\zeta_2^2
  - \mfrac{82032149}{34992}\zeta_3
  + \mfrac{5255296693}{839808}\zeta_2
  - \mfrac{2529781809149}{60466176}
 \Big)
  + \mathcal{O}(\pole{\ep})
\end{align}
\end{widetext}
A feature of our second method is that it provides
the results for all master integrals of a given family.
Often the simpler integrals with less lines could also be computed with the first approach, which moreover gave results through to transcendental
weight seven for almost all of the integrals.
This provided us with plenty of analytical cross checks.
For all integrals which were not checked by redundant analytical calculations,
we employed \texttt{Fiesta\;5}~\cite{Smirnov:2021rhf} to verify our analytical results
within a typical relative precision of $10^{-4}$ using a basis of finite integrals.


\bigskip
{\bf Results.}
Our calculation of the master integrals through to weight eight allows us to present
complete analytic results for $F_q$ and $F_g$. It is convenient to define their perturbative expansion in terms of the
bare strong coupling constant $\alpha_s^0$ as
\begin{align}
  \label{eq::FFbare}
  F_x &= 1 +
          \sum_{n\ge1} 
          \left(\frac{\alpha_s^0}{4\pi}\right)^n\!
          \left(\frac{4\pi}{e^{\gamma_E}}\right)^{n\epsilon}\!
          \left(\frac{\mu^2}{-q^2-i0} \right)^{n\epsilon}
          F_x^{(n)}\,,
\end{align}
with $x\in\{q,g\}$.
Here, $\gamma_E$ denotes Euler's constant,
and $\mu$ is the 't Hooft scale.

While the $\epsilon$ expansion of the fermionic corrections starts at order
$1/\epsilon^7$, the purely gluonic corrections also have $1/\epsilon^8$ poles and, correspondingly, zeta values with transcendental weight up to eight in the finite part.
Since all pole parts are known analytically from Ref.~\cite{Agarwal:2021zft},
see also \cite{Gracey:1994nn,Beneke:1995pq,Henn:2016men,Davies:2016jie,Grozin:2018vdn,Henn:2019rmi,Bruser:2019auj,Henn:2019swt,vonManteuffel:2020vjv,Moch:2017uml,Moch:2018wjh,Das:2019btv,Das:2020adl}, 
it is sufficient to consider the finite terms in the following.
The complete expressions can be found in a computer-readable ancillary
file attached to this Letter available on the arXiv.
We obtain for the finite part of the quark form factor
\begin{widetext}
{\scalefont{1.0}
\begin{align}
\label{eq:Fqres}
F_q^{(4)}\bigg|_{\epsilon^0} &= 
  \CS{ C_F ^4} \Big(
  - \mfrac{32384}{15}\zeta_{5,3}
  + \mfrac{739328}{45}\zeta_5 \zeta_3
  - \mfrac{66392}{27}\zeta_3^2 \zeta_2
  + \mfrac{7486576}{7875}\zeta_2^4
  - \mfrac{538913}{14}\zeta_7
  + \mfrac{20464}{5}\zeta_5 \zeta_2
  + \mfrac{1276}{15}\zeta_3 \zeta_2^2
  - \mfrac{192866}{27}\zeta_3^2
\nonumber\\& \hspace{-3.3ex}
  - \mfrac{737186}{63}\zeta_2^3
  - \mfrac{468509}{5}\zeta_5
  - \mfrac{269026}{9}\zeta_3 \zeta_2
  + \mfrac{304331}{30}\zeta_2^2
  + \mfrac{1823767}{48}\zeta_2
  + \mfrac{384343}{6}\zeta_3
  + \mfrac{6579473}{128}
 \Big) \nonumber\\& \hspace{-3.3ex}
+  \CS{ C_F^3 C_A} \Big(
  \mfrac{20948}{15}\zeta_{5,3}
  - \mfrac{18364}{5}\zeta_5 \zeta_3
  + \mfrac{3296}{9}\zeta_3^2 \zeta_2
  + \mfrac{5472536}{2625}\zeta_2^4
  + \mfrac{847073}{56}\zeta_7
  - \mfrac{445861}{45}\zeta_5 \zeta_2
  - \mfrac{1686022}{135}\zeta_3 \zeta_2^2
  - \mfrac{7935031}{162}\zeta_3^2
\nonumber\\& \hspace{-3.3ex}
  + \mfrac{73358647}{5670}\zeta_2^3
  + \mfrac{73074974}{405}\zeta_5
  + \mfrac{12562265}{162}\zeta_3 \zeta_2
  + \mfrac{11388811}{9720}\zeta_2^2
  - \mfrac{1176750697}{7776}\zeta_2
  + \mfrac{1623827681}{11664}\zeta_3
  - \mfrac{163952683523}{839808}
 \Big) \nonumber\\&  \hspace{-3.3ex}
+  \CS{ C_F^2 C_A^2} \Big(
   \mfrac{54254}{45}\zeta_{5,3}
  - \mfrac{45281}{9}\zeta_5 \zeta_3
  + \mfrac{38482}{27}\zeta_3^2 \zeta_2
  - \mfrac{21476059}{15750}\zeta_2^4
  - \mfrac{4032413}{144}\zeta_7
  + \mfrac{3166909}{270}\zeta_5 \zeta_2
  + \mfrac{4135486}{405}\zeta_3 \zeta_2^2
  + \mfrac{114329701}{1458}\zeta_3^2
\nonumber\\& \hspace{-3.3ex}
  - \mfrac{2153423}{420}\zeta_2^3
  - \mfrac{1049804117}{9720}\zeta_5
  - \mfrac{179904821}{2916}\zeta_3 \zeta_2
  - \mfrac{506374351}{29160}\zeta_2^2
  + \mfrac{38578372511}{209952}\zeta_2
  - \mfrac{33927798065}{104976}\zeta_3
  + \mfrac{1181337259783}{3779136}
 \Big) \nonumber\\&  \hspace{-3.3ex}
+  \CS{ C_F C_A^3 } \Big(
  - \mfrac{14161}{30}\zeta_{5,3}
  + \mfrac{21577}{6}\zeta_5 \zeta_3
  - \mfrac{1963}{3}\zeta_3^2 \zeta_2
  + \mfrac{10233079}{15750}\zeta_2^4
  + \mfrac{199283}{48}\zeta_7
  - \mfrac{20821}{9}\zeta_5 \zeta_2
  - \mfrac{68311}{45}\zeta_3 \zeta_2^2
  - \mfrac{3175501}{108}\zeta_3^2
\nonumber\\&  \hspace{-3.3ex}
  - \mfrac{565843}{1620}\zeta_2^3
  + \mfrac{75664147}{3240}\zeta_5
  + \mfrac{4325527}{324}\zeta_3 \zeta_2
  + \mfrac{1709477}{180}\zeta_2^2
  - \mfrac{842991647}{11664}\zeta_2
  + \mfrac{47856067}{324}\zeta_3
  - \mfrac{5465282473}{34992}
 \Big) \nonumber\\&  \hspace{-3.3ex}
+  \CS{\mfrac{d_F^{abcd}d_A^{abcd}}{N_F}} \Big(
  \mint{260}\zeta_{5,3}
  - \mint{5092}\zeta_5 \zeta_3
  - \mint{16}\zeta_3^2 \zeta_2
  - \mfrac{496766}{525}\zeta_2^4
  + \mint{3518}\zeta_7
  - \mfrac{4744}{3}\zeta_5 \zeta_2
  + \mfrac{6584}{15}\zeta_3 \zeta_2^2
  + \mfrac{39986}{9}\zeta_3^2
\nonumber\\& \hspace{-3.3ex}
  + \mfrac{526496}{945}\zeta_2^3
  - \mfrac{180566}{27}\zeta_5
  + \mfrac{3020}{3}\zeta_3 \zeta_2
  + \mfrac{1220}{9}\zeta_2^2
  + \mfrac{10570}{9}\zeta_2
  + \mfrac{169532}{27}\zeta_3
  - \mfrac{1580}{3}
 \Big) \nonumber\\&  \hspace{-3.3ex}

\mbox{+ contributions with closed fermion loop from Ref.~\cite{Lee:2021uqq}}
\\
\intertext{and for the finite part of the gluon form factor}
\label{eq:Fgres}
F_g^{(4)}\bigg|_{\epsilon^0} &= \CS{ C_A ^4} \Big(
  - \mfrac{2591}{90}\zeta_{5,3}
  + \mfrac{1018949}{90}\zeta_5 \zeta_3
  - \mfrac{35689}{27}\zeta_3^2 \zeta_2
  + \mfrac{18282694}{7875}\zeta_2^4
  - \mfrac{27705161}{504}\zeta_7
  + \mfrac{1160731}{270}\zeta_5 \zeta_2
  - \mfrac{1928564}{405}\zeta_3 \zeta_2^2
\nonumber\\& \hspace{-3.3ex}
  - \mfrac{1296845}{1458}\zeta_3^2
  - \mfrac{727183}{1134}\zeta_2^3
  + \mfrac{6161623}{243}\zeta_5
  - \mfrac{3233651}{729}\zeta_3 \zeta_2
  + \mfrac{54443689}{14580}\zeta_2^2
  + \mfrac{839716507}{104976}\zeta_2
  - \mfrac{84995881}{52488}\zeta_3
  + \mfrac{96887974603}{3779136}
 \Big) \nonumber\\& \hspace{-3.3ex}
+  \CS{\mfrac{d_A^{abcd}d_A^{abcd}}{N_A}} \Big(
   \mint{260}\zeta_{5,3}
  - \mint{5092}\zeta_5 \zeta_3
  - \mint{16}\zeta_3^2 \zeta_2
  - \mfrac{496766}{525}\zeta_2^4
  - \mfrac{6776}{3}\zeta_7
  - \mint{5016}\zeta_5 \zeta_2
  + \mfrac{2992}{3}\zeta_3 \zeta_2^2
  + \mfrac{31588}{3}\zeta_3^2
\nonumber\\&  \hspace{-3.3ex}
  + \mfrac{1073972}{945}\zeta_2^3
  - \mint{6460}\zeta_5
  + \mfrac{6752}{9}\zeta_3 \zeta_2
  + \mfrac{24616}{45}\zeta_2^2
  - \mfrac{4682}{27}\zeta_2
  - \mfrac{1310}{9}
  + \mfrac{68410}{9}\zeta_3
 \Big) \nonumber\\& \hspace{-3.3ex}

\mbox{+ contributions with closed fermion loop from Ref.~\cite{Lee:2021uqq}.}
\end{align}
}
\end{widetext}
We expressed our results in terms of invariants of a general
Lie algebra, where $C_R$ denotes the quadratic Casimir operator,
$d_R^{abcd}$ the fully symmetrical tensor originating from the
trace over generators,
and $N_R$ the dimension of the fundamental and adjoint
representation, $R=F,A$, respectively.
For a $SU(N_c)$ gauge group the invariants or color factors
are obtained as
\begin{align}
\label{eq:casimir}
C_F &= (N_c^2-1)/(2 N_c),\nonumber\\
C_A &= N_c,\nonumber\\
d_F^{abcd}d_A^{abcd} / N_F &= (N_c^2-1)(N_c^2+6)/48,\nonumber\\
d_A^{abcd}d_A^{abcd} / N_A &= N_c^2(N_c^2+36)/24.
\end{align}
All terms shown in Eqs.~\eqref{eq:Fqres} and \eqref{eq:Fgres} are new.
The complete four-loop results for $F_q$ and $F_g$ are obtained
after adding the fermionic contributions given in Eqs.~(10) and~(11)
of Ref.~\cite{Lee:2021uqq}.

We performed several checks of our results,
which we describe in the following.
First, the leading-color limit of Eq.~\eqref{eq:Fqres} agrees
with the result of Ref.~\cite{Lee:2016ixa}.
While all color structures of Eq.~\eqref{eq:Fqres} contribute in
this limit, it can be derived from just planar loop integrals, see
also Ref.~\cite{vonManteuffel:2019gpr} for an independent calculation.
Second, we observe that our weight-8 result for
$F_g^{(4)}/N_c^4$ agrees with the corresponding expression of the four-loop Sudakov form factor in ${\cal N}=4$ supersymmetric Yang Mills theory, see Eq.~(4.1) of
Ref.~\cite{Lee:2021lkc}, after expressing the color factors in terms of $N_c$ using Eqs.~\eqref{eq:casimir}.
Furthermore, after adjusting the QCD color factors such that the bosonic and
fermionic degrees of freedom are in the same color representation, \textit{i.e.}\
$C_F \to C_A, N_F \to N_A$ and
$d_A^{abcd} d_F^{abcd} \to d_A^{abcd} d_A^{abcd}$, we obtain identical results for the weight-8 coefficients of $F_q^{(4)}$ and $F_g^{(4)}$.  
These relations between the maximal transcendental parts involve all non-fermionic color coefficients of $F_q^{(4)}$ and
$F_g^{(4)}$ and test both leading and subleading color contributions.


\bigskip
{\bf Conclusions.}  In this Letter we provide the perturbative corrections
to the photon-quark and Higgs-gluon form factors at relative order $\alpha_s^4$.
This is the first complete calculation of vertex functions in four-loop massless QCD.
Our analytical results have been obtained by combining two
powerful multi-loop techniques: the direct integration of
finite master integrals and the method of differential equations.
The final expressions are presented in terms of zeta values with
transcendental weight up to eight, allowing for a straightforward
numerical evaluation.
Our results constitute the virtual contributions to a number of cross sections
and decay rates at N${}^4$LO, including Drell-Yan processes and gluon-fusion Higgs boson production at the LHC.



\bigskip
{\bf Acknowledgments.}
AvM and RMS gratefully acknowledge Erik Panzer for related collaborations.
This research was supported by the Deutsche Forschungsgemeinschaft (DFG,
German Research Foundation) under grant 396021762 — TRR 257 ``Particle Physics
Phenomenology after the Higgs Discovery'' and by the National Science
Foundation (NSF) under grant 2013859 ``Multi-loop amplitudes and precise
predictions for the LHC''.  The work of AVS and VAS was supported by the Ministry of Education and Science of the Russian Federation as part of the program of the Moscow Center for Fundamental and Applied Mathematics under agreement no. 075-15-2019-1621. The work of RNL is supported by the Russian Science Foundation, agreement no. 20-12-00205.
We acknowledge the High Performance Computing Center at Michigan State University for computing resources.  The Feynman diagrams were drawn with the help of {\tt
  Axodraw}~\cite{Vermaseren:1994je} and {\tt JaxoDraw}~\cite{Binosi:2003yf}.



\bibliography{ff_qcd}


\end{document}